\documentclass[12pt]{article}
\usepackage[dvips]{color}
\usepackage{epsfig}
\usepackage{amsmath}
\usepackage{multirow}
\usepackage{graphicx}
\def\Box{\hbox{$\rlap{$\sqcup$}\sqcap$}}
\begin{document}
\setcounter{page}{1}

\pagestyle{plain} \vspace{1cm}

\begin{center}
\Large{\bf Energy Conditions in $f(G)$ Modified Gravity with Non-minimal Coupling to Matter}\\
\small \vspace{1cm} {\bf A. Banijamali $^{a}$
\footnote{a.banijamali@nit.ac.ir}}, \quad {\bf B. Fazlpour $^{b}$
\footnote{b.fazlpour@umz.ac.ir}} and {\bf M. R. Setare$^{c}$
\footnote{rezakord@ipm.ir}}\\
\vspace{0.5cm}  $^{a}$ {\it Department of Basic Sciences, Babol
University of Technology, Babol, Iran\\} \vspace{0.5cm} $^{b}$
{\it Babol Branch, Islamic Azad University, Babol,
Iran\\}\vspace{0.5cm} $^{c}$  {\it Department of Science, Campus of Bijar, University of  Kurdistan  \\
Bijar, IRAN.}
\end{center}
\vspace{1.5cm}
\begin{abstract}
In this paper we study a model of modified gravity with non-minimal
coupling between a general function of the Gauss-Bonnet invariant,
$f(G)$, and matter Lagrangian from the point of view of the energy
conditions. Such model has been introduced in Ref. [21] for
description of early inflation and late-time cosmic acceleration. We
present the suitable energy conditions for the above mentioned model
and then, we use the estimated values of the Hubble, deceleration
and jerk parameters to apply the obtained energy conditions to the
specific class of modified Gauss-Bonnet models.
\end{abstract}
\newpage
\section{Introduction}
Nowadays it is strongly believed that the universe is experiencing
an accelerated expansion, and this is supported by many cosmological
observations [1-4]. This accelerated expansion can be explained in
terms of the so called dark energy ( for reviews see [5]) in the
framework of general relativity or by modification of general relativity. \\
The simplest type of modified gravity models is well known as $f(R)$
gravity where the Ricci scalar in the Einstein-Hilbert action is
replaced by a general function of the scalar curvature (see [6, 7] for reviews).\\
There are also other modified gravity models which are the
generalization of $f(R)$ gravity and among them, the modified
Gauss-Bonnet (GB) gravity {\it i.e.} $f(G)$ gravity, is more
interesting [8, 9]. The GB combination, $G$, is a topological
invariant in four dimensions, so in order to play some roles in the
field equations, one needs either couple GB term to a scalar field
like $f(\phi)G$, or choose $f(G)$ gravity where $f$ is an arbitrary function of $G$.\\
If we compare $f(G)$ gravity with other theories of modified gravity
we find some advantages in the Gauss-Bonnet gravity. For example in
the context of $f(G)$ gravity there exists a de-Sitter point that
can be used for cosmic acceleration [8, 9, 21]. Note that in $f(G)$
gravity, there are no problems [8, 9] with the Newton law,
instabilities and the anti-gravity regime. In comparing with the
simple $f(R)$ modified gravity we should mentioned that it is not
generally easy to construct viable $f(R)$ models that are consistent
with cosmological and local gravity constraints. The main reason for
this is that $f(R)$ gravity gives rise to a strong coupling between
DE and a non-relativistic matter in the Einstein frame [37]. However
there is no conformal transformation separating $G$ from scalar
field, unlike the $f(R)$ theory to obtain and Einstein frame action
with a canonical scalar field coupled to matter. Furthermore the
$f(G)$ models might be less constrained by local gravity constraints
relative to the $f(R)$ models. The main reason is that even in the
vacuum spherically symmetric background the Gauss-Bonnet scalar
takes a non-vanishing value. This property is different from $f(R)$
gravity in which the Ricci scalar $R$ vanishes in the vacuum
spherically symmetric background [38]. Moreover, In considering
alternative higher-order gravity theories, one is liable to be
motivated in pursuing models consistent and inspired by several
candidates of a fundamental theory of quantum gravity. Indeed,
motivations from string/M- theory predict that scalar field
couplings with the Gauss- Bonnet invariant, $G$, are important in
the appearance of non-singular early time cosmologies [33]. In
summary, modified $f(G)$ gravity represents a quite interesting
gravitational alternative to dark energy with more freedom if
compare with $f(R)$ gravity (for recent review see [14]). \\
The $f(G)$ gravity is an enrichment theory of modified gravity. A
number of its abilities are as follows: it has the possibility to
describe the inflationary era, a transition from a deceleration
phase to an acceleration phase, crossing the phantom divide line and
passing the solar system tests for a reasonable defined function $f$
[10-13]. \\
Moreover, a gravitational source of the inflation and dark energy
may be the non-minimal coupling of some geometrical invariants
function with matter Lagrangian. Such non-minimal modified gravity
has been introduced in Refs. [15, 16] for the study of gravity
assisted dark energy occurrence. It may be also applied for
realization of dynamical cancellation of cosmological constant [17].
The viability criteria for such theory was recently discussed in
Refs. [18-20]. Non- minimal coupling of $f(G)$ modified gravity with
matter Lagrangian has been investigated in [21]. It is shown [21],
that such a model can easily unify the early inflation with
late-time cosmic acceleration for the special choise of
gravitational functions. Clearly, as any model in $f(G)$ theory,
there are particular conditions which have to be satisfied in order
to ensure that the model is viable and physically meaningful
[22-24]. In the present work we study the above mentioned model of
$f(G)$ gravity with non-minimal coupling to matter from the
viewpoint of the energy conditions.\\
Furthermore, the energy conditions are essential for studying the
singularity theorems as well as the theorems of black hole
thermodynamics. For example the Hawking-Penrose singularity theorems
invoke the weak and strong energy conditions, whereas the proof of
the second law of black hole thermodynamics needs the null energy
condition [25]. Various forms of energy conditions namely, strong,
weak, dominant and null energy conditions are obtained using
Raychaudhuri equation along with attractiveness property of gravity
[25, 26]. Energy conditions have been widely studied in the
literature. Phantom field potential [27], expansion history of the
universe [28] and evolution of the deceleration parameter [29] have
been investigated using classical energy conditions of general
relativity. Energy conditions in the context of $f(R)$ gravity have
been derived in [30] and following the formalism developed in [30],
several authors have studied some cosmological issues in modified
$f(R)$ gravity from this point of view (see for example [31, 32]).
Also, appropriate energy conditions for the $f(R)$ gravity
non-minimally coupled with matter Lagrangian has been investigated in [22]. \\
In a recent paper [33], the general energy conditions for modified
Gauss-Bonnet gravity or $f(G)$ gravity have been presented and
viability of specific realistic forms of $f(G)$ proposed in [21]
have been analyzed by imposing the weak energy condition.\\
In this paper we generalize procedure developed in [33] for $f(G)$
theories to the gravity model with non-minimal coupling of $f(G)$ to
the matter. Then, we use the estimated values of the Hubble,
deceleration and jerk parameters to apply the obtained energy
conditions to the specific class of models.\\
An outline of this paper is as follows: in the next section we
review the field equations of modified $f(G)$ theory non-minimally
coupled with matter. In section 3 we present the suitable energy
conditions for such a model. In section 4 we examine the weak energy
condition for two specific class of $f(G)$ models. Section 5 is
devoted to our conclusions.\\
\section{Field Equations}
Our starting action for modified Gauss-Bonnet gravity non-minimally
coupled with matter is as follows [21]:
\begin{equation}
S=\int
d^4x\sqrt{-g}\big[\frac{1}{2\kappa^2}R+f(G)\mathcal{L}_{m}\big],
\end{equation}
where $g$ is the determinant of the metric tensor $g_{\mu\nu}$, $R$
is the Ricci scalar and $f(G)$ is a general function of Gauss-Bonnet
invariant $G$. One notes that there is a non-minimal coupling
between $f(G)$ gravity and the matter
Lagrangian $\mathcal{L}_m$ in (1).\\
The theory of kind (1) has been previously investigated in order to
unify inflation and late-time cosmic acceleration [21].\\
The Gauss-Bonnet term is given by,
\begin{equation}
G=R^{2}-4R_{\mu\nu}R^{\mu\nu}+R_{\mu\nu\rho\sigma}
R^{\mu\nu\rho\sigma}.
\end{equation}
Varying the action (1) with respect to $g_{\mu\nu}$ leads to:
\begin{eqnarray}
0&=&\frac{1}{2\kappa^{2}}\big(R_{\mu\nu}-\frac{1}{2}g_{\mu\nu}R\big)+2f'(G)\mathcal{L}_{m}R\,R_{\mu\nu}\nonumber\\
&-&4f'(G)\mathcal{L}_{m}R_{\mu}\,^{\rho}R_{\nu\rho}+2f'(G)\mathcal{L}_{m}R_{\mu\rho\sigma\lambda}R_{\nu}\,^{\rho\sigma\lambda}
+4f'(G)\mathcal{L}_{m}R_{\mu\rho\sigma\nu}R^{\rho\sigma}\nonumber\\
&+&2\Big(\big(g_{\mu\nu}\Box-\nabla_{\mu}\nabla_{\nu}\big)
f'(G)\mathcal{L}_{m}\Big)R+4\big(\nabla^{\rho}\nabla_{\mu}
f'(G)\mathcal{L}_{m}\big)R_{\nu\rho}+4\big(\nabla^{\rho}\nabla_{\nu}
f'(G)\mathcal{L}_{m}\big)R_{\mu\rho}\nonumber\\
&+&4\big(\Box
f'(G)\mathcal{L}_{m}\big)R_{\mu\nu}-4\big(g_{\mu\nu}\nabla^{\lambda}
\nabla^{\rho}f'(G)\mathcal{L}_{m}\big)
R_{\lambda\rho}+4\big(\nabla^{\rho}
\nabla^{\lambda}f'(G)\mathcal{L}_{m}\big)R_{\mu\rho\nu\lambda}+f(G)T_{\mu\nu},\nonumber\\
\end{eqnarray}
where,
\begin{eqnarray}
T_{\mu\nu}=\frac{2}{\sqrt{-g}}\frac{\delta}{\delta
g^{\mu\nu}}\Big(\int d^{4}x\sqrt{-g}\mathcal{L}_{m}\Big),
\end{eqnarray}
is the energy-momentum tensor of the ordinary matter and prime
stands for derivative with respect to $G$. For a flat
Friedman-Robertson-Walker (FRW) metric with scale factor $a(t)$,
\begin{eqnarray}
ds^{2}=-dt^{2}+a^{2}(t)(dr^{2}+r^{2}d\Omega^{2}),
\end{eqnarray}
and assuming matter as a perfect fluid, the field equation (3) has
the following simple forms,
\begin{eqnarray}
&& \hspace{-5mm}
\frac{3}{\kappa^{2}}H^{2}+24H^{3}\frac{d}{dt}\big(f'(G)\mathcal{L}_{m}\big)-G
f'(G)\mathcal{L}_{m}=f(G)\rho,\\
&& \hspace{-5mm}
\frac{1}{\kappa^{2}}\big(2\dot{H}+3H^{2}\big)+8H^{2}\frac{d^{2}}{dt^{2}}\big({f}'(G)\mathcal{L}_{m}\big)+16H\frac{d}{dt}\big({f}'(G)\mathcal{L}_{m}\big)
(\dot{H}+H^{2})-G f'(G)\mathcal{L}_{m}=-f(G)p,\nonumber\\
\end{eqnarray}
where $\rho$ and $p$ are the energy density and pressure,
respectively. The overdot denotes a time derivative.\\
In order to write the field equations as those in general
relativity, one can define the effective energy density and
pressure,
\begin{equation}
\rho_{eff}=\frac{3}{\kappa^{2}}H^{2},
\end{equation}
and
\begin{equation}
p_{eff}=-\frac{1}{\kappa^{2}} \big( 2\dot H+3H^{2} \big),
\end{equation}
where $\rho_{eff}$ and $p_{eff}$ are the effective energy density
and pressure, respectively and given by,
\begin{equation}
\rho_{eff}=f(G)\Big[G
\frac{f'(G)}{f(G)}\mathcal{L}_{m}-24H^{3}\dot{G}\frac{f''(G)}{f(G)}\mathcal{L}_{m}+\rho\Big],
\end{equation}
and
\begin{equation}
p_{eff} = f(G)\Bigg[-G
\frac{f'(G)}{f(G)}\mathcal{L}_{m}+8H^{2}\Big(\ddot{G}\,\frac{f''(G)}{f(G)}+\dot{G}^{2}\,\frac{f'''(G)}{f(G)}\Big)\mathcal{L}_{m}
+16H\dot{G}\,\frac{f''(G)}{f(G)}\big(\dot{H}+H^{2}\big)\mathcal{L}_{m}+p\Bigg],
\end{equation}
where, we have used the equations (6) and (7).\\

\section{Energy conditions}
The energy conditions originate from the Raychaudhury equation
together with the attractiveness of the gravity [26]. The
Raychaudhury equation in the case of a congruence of null geodesics
defined by the vector field $k^{\mu}$ is given by,
\begin{equation}
\frac{d\theta}{d\tau}=-\frac{1}{2}\theta^{2}-\sigma_{\mu\nu}\sigma^{\mu\nu}+\omega_{\mu\nu}
\omega^{\mu\nu}-R_{\mu\nu}k^{\mu}k^{\nu},
\end{equation}
where $R_{\mu\nu}$ is the Ricci tensor, $\theta$ is the expansion
parameter, $\sigma_{\mu\nu}$ and $\omega_{\mu\nu}$ are the shear and
the rotation associate to the congruence respectively.\\
For any hypersurface of orthogonal congruences
($\omega_{\mu\nu}=0$), the condition of attractiveness of gravity
($\frac{d\theta}{d\tau}<0$) yields to
$R_{\mu\nu}k^{\mu}k^{\nu}\geq0$
($\sigma_{\mu\nu}\sigma^{\mu\nu}\geq 0$).\\
This condition in the context of general relativity expresses by
$T_{\mu\nu}k^{\mu}k^{\nu}\geq 0$.\\
The Raychaudhury equation satisfies for any geometrical theory of
gravitation. In the other hand the theory of kind (1) has an
Einstein-Hilbert term and evaluating $R_{\mu\nu}k^{\mu}k^{\nu}$ is
straightforward. So, using the effective gravitation field
equations, the energy conditions in the case of our model are given
by,
\begin{equation}
\rho_{eff}+p_{eff}\geq 0 \,\,\,\,\,\, for\,\, null\,\, energy\,\,
condition \,\,(NEC),
\end{equation}
\vspace{0.03cm}
\begin{equation}
\rho_{eff}\geq 0 \,\, and\,\,\rho_{eff}+p_{eff}\geq 0 \,\,\,\,\,\,
for\,\, weak\,\, energy\, \,condition\,\, (WEC),
\end{equation}
\vspace{0.05cm}
\begin{equation}
\rho_{eff}+3p_{eff}\geq 0\,\,and\,\,\rho_{eff}+p_{eff}\geq 0
\,\,\,\,\,\, for\,\, strong\, \,energy \,\,condition\, \,(SEC),
\end{equation}
and
\begin{equation}
\rho_{eff}\geq 0\,\, and\,\,\rho_{eff}\pm p_{eff}\geq 0 \,\,\,\,\,\,
for\,\, dominant\,\, energy \,\,condition \,\,(DEC).
\end{equation}
Inserting equations (10) and (11) into equations (13)-(16) lead to
the following respective forms,

$$NEC:\rho+p
-8H^{3}\dot{G}\frac{f''(G)}{f(G)}\mathcal{L}_{m}+8H^{2}\Big(\ddot{G}\,\frac{f''(G)}{f(G)}+
\dot{G}^{2}\,\frac{f'''(G)}{f(G)}\Big)\mathcal{L}_{m}$$
\begin{eqnarray}
+16H\dot{H}\dot{G}\,\frac{f''(G)}{f(G)}\mathcal{L}_{m}\geq0\,,
\end{eqnarray}
\begin{equation}
WEC:\rho+G\frac{f'(G)}{f(G)}\mathcal{L}_{m}-24H^{3}\dot{G}\frac{f''(G)}{f(G)}\mathcal{L}_{m}\geq0\,,
\,\,\,\,\,\,\,\,\rho_{eff}+p_{eff}\geq 0.
\end{equation}

$$SEC:\rho+3p-2G\frac{f'(G)}{f(G)}\mathcal{L}_{m}+24H^{3}\dot{G}\frac{f''(G)}{f(G)}\mathcal{L}_{m}+24H^{2}
\Big(\ddot{G}\,\frac{f''(G)}{f(G)}+\dot{G}^{2}\,\frac{f'''(G)}{f(G)}\Big)\mathcal{L}_{m}
$$
\begin{equation}
+48H\dot{H}\dot{G}\,\frac{f''(G)}{f(G)}\mathcal{L}_{m}\geq0\,,
\,\,\,\,\,\,\,\,\rho_{eff}+p_{eff}\geq 0\,.
\end{equation}

$$DEC:\rho-p+2G\frac{f'(G)}{f(G)}\mathcal{L}_{m}-40H^{3}\dot{G}\frac{f''(G)}{f(G)}\mathcal{L}_{m}-8H^{2}
\Big(\ddot{G}\,\frac{f''(G)}{f(G)}+\dot{G}^{2}\,\frac{f'''(G)}{f(G)}\Big)\mathcal{L}_{m}
$$
\begin{equation}
-16H\dot{H}\dot{G}\,\frac{f''(G)}{f(G)}\mathcal{L}_{m}\geq0\,,
\,\,\,\,\,\,\,\,\rho_{eff}+p_{eff}\geq
0\,,\,\,\,\,\,\,\,\,\,\rho_{eff}\geq 0\,.
\end{equation}
In order to analyze the model of type (1) with non-minimal coupling
between $f(G)$ gravity and matter from the point of view of energy
conditions, we use the standard terminology in studying energy
conditions for modified gravity theories. To this end, according to
the Hubble parameter $H=\frac{\dot{a}}{a}$ we define the
deceleration ($q$), jerk ($j$), and snap ($s$) parameters as,
\begin{equation}
q=-\frac{1}{H^{2}}\frac{\ddot{a}}{a}\;, \qquad
j=\frac{1}{H^{3}}\frac{\dddot{a}}{a}\;, \quad {\rm{and}} \quad
s=\frac{1}{H^{4}}\frac{\ddddot{a}}{a}\;.
\end{equation}
In terms of above parameters, time derivatives of the Hubble
parameter can be expressed as,
\begin{eqnarray}
&\dot{H}=-H^{2}(1+q)\;, \\
&\ddot{H}=H^{3}(j+3q+2)\;, \\
&\dddot{H}=H^{4}(s-2j-5q-3)\;.
\end{eqnarray}
The Gauss-Bonnet invariant in FRW background is as $G=24H^{2} \big(
H^{2}+\dot H \big)$. So, $G$ and its time derivatives can be written
as,
\begin{eqnarray}
&G=-24H^{4} q\;, \\
&\dot{G}=24H^{5}\big(2q^{2}+3q+j\big)\;, \\
&\ddot{G}=-24H^{6}\big(2q^{3}+22q^{2}+6qj+15q+4j-s-3\big)\;.
\end{eqnarray}
By inserting relations (22)-(27) into equations (17)-(20) one can
obtain the following forms of energy conditions

$$\rho_{0}+p_{0}-192H_{0}^{8}\big(5q_{0}^{3}+34q_{0}^{2}+8q_{0}j_{0}+24q_{0}+7j_{0}-s_{0}-3\big)
\frac{f_{0}''(G)}{f_{0}(G)}\mathcal{L}_{m}$$
\begin{equation}
+4608H_{0}^{12}\big(2q_{0}^{2}+3q_{0}+j_{0}\big)^{2}
\frac{f_{0}'''(G)}{f_{0}(G)}\mathcal{L}_{m}\geq0,
\end{equation}
for null,
\begin{equation}
\rho_{0}-24H_{0}^{4}q_{0}\frac{f_{0}'(G)}{f_{0}(G)}\mathcal{L}_{m}-576H_{0}^{8}
\big(2q_{0}^{2}+3q_{0}+j_{0}\big)\frac{f_{0}''(G)}{f_{0}(G)}\mathcal{L}_{m}\geq0,
\end{equation}
for weak,
$$\rho_{0}+3p_{0}+48H_{0}^{4}q_{0}\frac{f_{0}'(G)}{f_{0}(G)}\mathcal{L}_{m}-576H_{0}^{8}\big(6q_{0}
^{3}+30q_{0}^{2}+8q_{0}j_{0}+18q_{0}+5j_{0}-s_{0}-3\big)\frac{f_{0}''(G)}{f_{0}(G)}\mathcal{L}_{m}$$
\begin{equation}
+13824H_{0}^{12}\big(2q_{0}^{2}+3q_{0}+j_{0}\big)^{2}
\frac{f_{0}'''(G)}{f_{0}(G)}\mathcal{L}_{m}\geq0,
\end{equation}
for strong and
$$\rho_{0}-p_{0}-48H_{0}^{4}q_{0}\frac{f_{0}'(G)}{f_{0}(G)}\mathcal{L}_{m}+192H_{0}^{8}\big(6q_{0}
^{3}+22q_{0}^{2}+8q_{0}j_{0}+6q_{0}+j_{0}-s_{0}-3\big)\frac{f_{0}''(G)}{f_{0}(G)}\mathcal{L}_{m}$$
\begin{equation}
-4608H_{0}^{12}\big(2q_{0}^{2}+3q_{0}+j_{0}\big)^{2}
\frac{f_{0}'''(G)}{f_{0}(G)}\mathcal{L}_{m}\geq0,
\end{equation}
for dominant energy condition respectively. The subscript $0$ stands
for the present value of quantities. These forms of energy
conditions are suitable to impose bounds on a given $f(G)$ model by
using the estimate values of the $q_{0}, j_{0}$ and $s_{0}$.
\section{Specific models}
Now, let us see how the above energy conditions work for specific
type of $f(G)$ models. In what follows we focus just on WEC
inequality (29). Our reason apart from the simplicity is that all
other energy conditions depend on the present value of snap
parameter $s_{0}$ and as mentioned in [30] until now no reliable
measurement of this parameter has been reported. Also, for
simplicity we only examine the case in which $p = \rho = 0$,
although this is not a physically interesting case, but this is
easily corrected, since one can always add a positive energy density
or pressure from matter and/or radiation to any model satisfying the
WEC , and it will still satisfy the WEC [33].\\
\, \,

\textbf{Case 1:} Our first example is coming from Ref. [9],
\begin{equation}
f(G)=\alpha \,G^{\,n}.
\end{equation}
It has been shown in [9] that this type of $f(G)$ theory can produce
quintessence, phantom or cosmological constant cosmology and has the
possibility of realizing transition from the deceleration to the
acceleration era. It was shown in [39] that the model of type (32)
with $n<0$, is not cosmologically viable because of separatrices
between radiation and dark energy dominations. In [38] it was found
that the model (32) with $n>0$, can be consistent with solar
system tests for $n\leq0.074$ if the Gauss-Bonnet term is responsible for dark energy.\\
Inserting $f(G)$ from (32) and their derivatives into equation (29)
for WEC yields to
\begin{equation}
(-1)^{n}\alpha \,[A n^{2}-(A+1)n]\mathcal{L}_{m}\leq0,
\end{equation}
where $A=\frac{2q^{2}+3q+j}{q^{2}}$.\\
The following estimated values of deceleration and jerk parameters
[34, 35]; $q_{0}=-8.81\pm0.14$ and $j_{0}=2.16^{+0.81}_{-.75}$ allow
us to further analyze equation (33). Note that the roots of $[A
n^{2}-(A+1)n]=0$ are $(0,1+\frac{1}{A})$, so it has positive values
for $0>n>1.63$ and negative values for $0\leq n\leq 1.63$. In
addition, the condition required for attractiveness of gravity {\it
i.e.} $f(G)>0$  leads to $\alpha(-1)^{n}\big(2.09H\big)^{4n}>0$.
Now, using the above explanations, one can study the inequality
(33) as follows:\\
\textbf{(i)} for $\alpha>0$, one concludes from the attractivness
property of gravity that $n$ should be even and then the equation
(33) tell us that $0\leq
n\leq 1.63$, and therefore there is no allowed value for $n$.\\
\textbf{(ii)} for $\alpha<0$, $n$ should be odd and equation (33)
leads to $0>n>1.63$ so the allowed sets of $n$s are $n=\{3, 5,
7,...\}$ and $n=\{-1, -3, -5,...\}$.\\

\textbf{Case 2:} In the second example we consider a class of models
as follows:
\begin{equation}
f(G)=\alpha G^{\,n}\big(1+\beta G^{m}\big).
\end{equation}
The models of kind (34) have been proposed in Ref. [21] in order to
address the late-time cosmic acceleration. It has been shown in [40]
that the model (34) is consistent with local tests and cosmological
bounds. The typical property of such theory is the presence of the
effective cosmological constant epochs in such a way that early-time
inflation and late-time cosmic acceleration are naturally unified
within single model. It is shown that classical instability does not
appear here and Newton law is respected. Some discussion of possible
anti-gravity regime appearance and related modification of the
theory is also done in [40]. Furthermore, it is shown in [36] that
four types of future singularities can be cured in such a models if
one considers $n>0$ and $m<0$.\\
By inserting the above $f(G)$  and their derivatives into equation
(29), WEC reads
\begin{equation}
\Big[\alpha (-1)^{n} a_{n}
+\alpha\beta(-1)^{n+m}(a_{n}+a_{m}+2Amn)\big(24H^{4}q\big)^{m}\Big]
\mathcal{L}_{m}\leq0,
\end{equation}
where $a_{n}=A n^{2}-(A+1)n$ and $a_{n}+a_{m}+2Amn=A
(n+m)^{2}-(A+1)(n+m)$. For the next purposes, we mention the roots
of $a_{n}=0$ and $a_{n}+a_{m}+2Amn=0$ are $n=(0,1+\frac{1}{A})$ and
$n+m=(0,1+\frac{1}{A})$ respectively. So, they have positive values
for $0>n>1.63$ and $0>n+m>1.63$ and negative values for $0\leq n\leq
1.63$ and $0\leq n+m\leq 1.63$ respectively. In addition, the
condition $f(G)>0$ leads to
$\alpha(-1)^{n}\big(2.09H\big)^{4n}\Big(1+\beta(-1)^{m}\big(2.09H\big)
^{4m}\Big)>0$.\\
Now, by using the above necessary condition one can study the
inequality (35) as the
following cases:\\
\\
\textbf{(i)} In the first situation we assume that $a_{n}>0$ and
$a_{n}+a_{m}+2Amn>0$
or $a_{n}<0$ and $a_{n}+a_{m}+2Amn<0$ so we have the following possibilities:\\
\\
\textbf{(i1)}\,\, for $\alpha(-1)^{n}>0$, \,\,if
$|\frac{a_{n}}{a_{n}+a_{m}+2Amn}|\big(2.09H\big) ^{-4m}\leq \beta <
\big(2.09H\big) ^{-4m}$ then $m$ should be odd and if
$-\big(2.09H\big) ^{-4m}< \beta
<-|\frac{a_{n}}{a_{n}+a_{m}+2Amn}|\big(2.09H\big) ^{-4m} $ then $m$
should be even.\\
\textbf{(i2)}\,\, for $\alpha(-1)^{n}<0$, \,\, it is impossible
to realize the conditions (35) and $f(G)>0$ simultaneously.\\
\\\\
\textbf{(ii)} Here, in the second situation we  present the
conditions required for WEC fulfilment for $a_{n}<0$ and
$a_{n}+a_{m}+2Amn>0$
or $a_{n}>0$ and $a_{n}+a_{m}+2Amn<0$:\\
\\
\textbf{(ii1)}\,\, for $\alpha(-1)^{n}>0$, \,\,if $\beta \geq
|\frac{a_{n}}{a_{n}+a_{m}+2Amn}|\big(2.09H\big) ^{-4m}$ then $m$
should be even and if $\beta \leq
-|\frac{a_{n}}{a_{n}+a_{m}+2Amn}|\big(2.09H\big) ^{-4m}$ then $m$
should be odd.\\
\textbf{(ii2)}\,\, for $\alpha(-1)^{n}<0$, \,\, if
$\beta>\big(2.09H\big) ^{-4m}$ then $m$ should be odd and if
$\beta<-\big(2.09H\big) ^{-4m}$ then $m$ should be even.\\

\section{Conclusion}
In this work we investigated a model of modified gravity with
non-minimal coupling between modified Gauss-Bonnet gravity, $f(G)$,
and matter Lagrangian described by action (1) from the viewpoint of
the energy conditions. We derived the suitable energy
conditions inequalities for such a model as equations (17)- (20).\\
We examined the WEC inequality equation (29) for two class of viable
models of $f(G)$ gravity presented in equations (32) and (34). We
concluded that the model $f(G)=\alpha \,G^{\,n}$ obey the WEC only
for $\alpha<0$. For the model $f(G)=\alpha G^{\,n}\big(1+\beta
G^{m}\big)$ our result summarized in subcases \textbf{(i1)},
\textbf{(i2)}, \textbf{(ii1)} and \textbf{(ii2)}. One can use our
method and in particular equations (28)-(31) to study the physical
implications of any $f(G)$ model with non-minimal coupling to
matter.\\
Finally, we mention the following important point: as emphasized in
[30], although the energy conditions in modified gravity theories
have well-founded physical motivation (the Raychaudhury equation
together with the attractiveness property of gravity) the question
as to whether they should be applied to any modified gravity theory
is an open question which is ultimately related to the confrontation
between theory and observations.\\


\begin{thebibliography}{11}
\bibitem{p1}
S. Perlmutter, \emph{ et al.},  Supernova Cosmology Project
Collaboration, \emph{ Astrophys. J.} \textbf{517}, 565 (1999).
\bibitem{p2}
C. L. Bennett, \emph{ et al.}, \emph{ Astrophys. J. Suppl.}
\textbf{148}, 1 (2003).
\bibitem{p3}
M. Tegmark, \emph{ et al.}, SDSS Collaboration, \emph{ Phys. Rev. D}
\textbf{69}, 103501 (2004).
\bibitem{p4}
S.W. Allen, \emph{ et al.}, \emph{ Mon. Not. R. Astron. Soc.}
\textbf{353}, 457 (2004).
\bibitem{p5}
E. J. Copeland, M. Sami, S. Tsujikawa, \emph{Int. J. Mod. Phys. D}
\textbf{15}, 1753 (2006); Y. F. Cai, E. N. Saridakis, M. R. Setare,
J. Q. Xia, \emph{Physics Reports} \textbf{493}, 1 (2010); M. Li, X.
D. Li, S. Wang, Y. Wang, arXiv:1103.5870 [astro-ph.CO].
\bibitem{p6}
S. Nojiri and S. D. Odintsov, \emph{Int. J. Geom. Math. Mod. Phys.}
\textbf{4}, 115 (2007).
\bibitem{p7}
S. Nojiri and S. D. Odintsov, arXiv:0801.4843 [astro-ph];
arXiv:0807.0685 [hep-th]; T. P. Sotiriou and V. Faraoni, \emph{Rev.
Mod. Phys.} \textbf{82}, 451 (2010); F. S. N. Lobo, arXiv:0807.1640
[gr-qc] and S. Capozziello and M. Francaviglia,\emph{ Gen. Rel.
Grav.} \textbf{40}, 357 (2008).
\bibitem{p8}
S. Nojiri and S. D. Odintsov, \emph{Phys. Lett.} B\textbf{ 631}, 1
(2005).
\bibitem{p9}
G. Cognola, E. Elizalde, S. Nojiri, S. D. Odintsov and S. Zerbini,
\emph{Phys. Rev. D} \textbf{75}, 086002 (2007).
\bibitem{p10}
S. Nojiri, S. D. Odintsov, S. Ogushi, \emph{Int. J. Mod. Phys. A}
\textbf{17}, 4809 (2002); B. M. Leith, I. P. Neupane, \emph{J.
Cosmol. Astropart. Phys.} \textbf{0705}, 019 (2007).
\bibitem{p11}
J. Sadeghi, M. R. Setare and A. Banijamali, \emph{Phys. Lett. B}
\textbf{679}, 302 (2009); J. Sadeghi, M. R. Setare and A.
Banijamali, \emph{Eur. Phys. J. C} \textbf{64}, 433 (2009); M. R.
Setare and E. N. Saridakis , \emph{Phys. Lett. B} \textbf{670}, 1
(2008); M. R. Setare, \emph{Int. J. Mod. Phys. D} \textbf{17}, 2219
(2008); M. R. Setare, \emph{Chin. Phys. Lett.} \textbf{26}, 029501
(2009).
\bibitem{p12}
A. De Felice, S. Tsujikawa, \emph{Phys. Lett. B} \textbf{675}, 1
(2009).
\bibitem{p13}
A. De Felice, S. Tsujikawa, \emph{Phys. Rev. D} \textbf{80}, 063516
(2009).
\bibitem{p14}
S. Nojiri and S. D. Odintsov, arXiv:1011.0544 [gr-qc]; S.
Capozziello, M. D. Laurentis, arXiv:1108.6266v2 [gr-qc].

\bibitem{p15}
S. Nojiri and S. D. Odintsov, {\it Phys. Lett. B} {\bf 599}, 137
(2004); PoS {\bf WC2004}, 024 (2004) arXiv:0412030[hep-th].
\bibitem{p16}
G. Allemandi, A. Borowiec, M. Francaviglia, and S. D. Odintsov, {\it
Phys. Rev. D} {\bf 72}, 063505 (2005).
\bibitem{p17}
S. Mukohyama and L. Randall, {\it Phys. Rev. Lett.} {\bf 92}, 211302
(2004); T. Inagaki, S. Nojiri and S. D. Odintsov, {\it JCAP} {\bf
0506}, 010 (2005) [arXiv:gr-qc/0504054]; A. D. Dolgov and M.
Kawasaki, arXiv:0307442[astro-ph].
\bibitem{p18}
O. Bertolami, C. Boehmer, T. Harko, and F. Lobo, {\it Phys. Rev. D}
{\bf 75}, 104016 (2007); T. Koivisto, {\it Class. Quant. Grav.} {\bf
23}, 4289 (2006).
\bibitem{p19}
V. Faraoni, {\it Phys. Rev. D} {\bf 76}, 127501 (2007).
\bibitem{p20}
O. Bertolami and J. Paramos, {\it Phys. Rev. D} {\bf 77}, 084018
(2008).
\bibitem{p21}
S. Nojiri, S. D. Odintsov, and P. Tretyakov, {\it Phys. Lett. B}
{\bf 651}, 224 (2007); S. Nojiri, S. D. Odintsov and P. V.
Tretyakov, \emph{Prog. Theor. Phys. Supp} \textbf{172}, 81 (2008).
\bibitem{p22}
O. Bertolami and M. C. Sequeira,\emph{ Phys. Rev. D} \textbf{79},
104010 (2009).
\bibitem{p23}
J. H. Kung, \emph{Phys. Rev. D} \textbf{52}, 6922 (1995) ;
\emph{Phys. Rev. D} \textbf{53}, 3017 (1996).
\bibitem{p24}
S. E. Perez Bergliaffa, \emph{Phys. Lett.} B \textbf{642}, 311
(2006).
\bibitem{p25}
S. W. Hawking and G. F. R. Ellis, \emph{The Large Scale Structure of
Spacetime} (Cambridge University Press, England, 1973); R. M. Wald,
\emph{General Relativity} (University of Chicago Press, Chicago,
1984).
\bibitem{p26}
S. Carroll, \emph{Spacetime and Geometry: An Introduction to General
Relativity,} (Addison Wesley, New York, 2004).
\bibitem{p27}
J. Santos and J. S. Alcaniz, \emph{Phys. Lett. B} \textbf{619}, 11
(2005).
\bibitem{p28}
M. Visser, \emph{Science} \textbf{276}, 88 (1997); \emph{Phys. Rev.
D} \textbf{56}, 7578 (1997) ; J. Santos, J. S. Alcaniz and M. J.
Rebou¸cas, \emph{Phys. Rev. D} \textbf{74}, 067301 (2006); J.
Santos, J. S. Alcaniz, N. Pires and M. J. Rebou¸cas, \emph{Phys.
Rev. D }\textbf{75}, 083523 (2007); A. A. Sen and R. J. Scherrer,
\emph{Phys. Lett. B} \textbf{659}, 457 (2008); J. Santos, J. S.
Alcaniz, M. J. Rebou¸cas and N. Pires, \emph{Phys. Rev. D}
\textbf{76}, 043519 (2007).
\bibitem{p29}
Y. G. Gong, A. Wang, Q. Wu and Y. Z. Zhang, {\it JCAP} {\bf 0708},
018 (2007); Y. Gong and A. Wang, arXiv:0705.0996v1 [astro-ph].
\bibitem{p30}
J. Santos, J. S. Alcaniz, M. J. Reboucas and F. C. Carvalho,
\emph{Phys. Rev. D} \textbf{76}, 083513 (2007).
\bibitem{p31}
J. Santos, M. J. Rebou¸cas and J. S. Alcaniz, \emph{Int. J. Mod.
Phys. D} \textbf{19}, 1315 (2010).
\bibitem{p32}
K. Atazadeh, A. Khaleghi, H. R. Sepangi and Y. Tavakoli, \emph{Int.
J. Mod. Phys. D} \textbf{18}, 1101 (2009).
\bibitem{p33}
N. M. Garcia, T. Harko, Francisco S. N. Lobo and Jos´e P. Mimoso,
\emph{Phys. Rev. D} \textbf{83}, 104032 (2011).
\bibitem{p34}
D. Rapetti, S. W. Allen, M. A. Amin and R.D. Blandford, \emph{Mont.
Not. R. Soc}. \textbf{375}, 1510 (2007).
\bibitem{p35}
N. J. Poplawski,\emph{ Class. Quant. Grav.} \textbf{24}, 3013
(2007).
\bibitem{p36}
K. Bamba, S. D. Odintsov, L. Sebastiani and S. Zerbini, \emph{Eur.
Phys. J. C} \textbf{67}, 295 (2010).
\bibitem{p37}
L. Amendola, D. Polarski and S. Tsujikawa,\emph{Phys. Rev. Lett}
\textbf{98}, 131302 (2007).
\bibitem{p38}
S. C. Davis, arXiv:0709.4453 [hep-th].
\bibitem{p39}
A. De Felice and M. Hindmarsh,\emph{JCAP} \textbf{0706}, 028 (2007).
\bibitem{p40}
S. Nojiri and S. D. Odintsov,\emph{Phys. Lett. B} \textbf{657}, 238
(2007).
\end{thebibliography}
\end{document}